\documentclass[12pt]{iopart}
\usepackage{mathrsfs}
\usepackage{graphicx}
\usepackage{graphics}
\usepackage{epsfig}
\usepackage{color}

\begin{document}

\title[Zadorosny \emph{et al.}]{Dynamics and heat diffusion of Abrikosov's
	vortex-antivortex pairs during an annihilation process}

\author{E C S Duarte$^1$, E Sardella$^2$, W A Ortiz$^3$
	and R Zadorosny$^1$}

\address{$^1$Departamento de F\'isica e Qu\'imica,
	Univiversidade Estadual Paulista (UNESP), Faculdade
	de Engenharia de Ilha Solteira, Caixa Postal 31,
	15385-000 Ilha Solteira-SP, Brazil}
\address{$^2$Departamento de F\'isica, Univiversidade
	Estadual Paulista (UNESP), Faculdade de Ci\^encias,
	Caixa Postal 473, 17033-360, Bauru-SP, Brazil }
\address{$^3$Departamento de F\'isica, Universidade
	Federal de S\~ao Carlos - UFSCar, 13565-905,
	S\~ao Carlos-SP, Brazil}

\ead{rafazad@gmail.com}

\begin{abstract}

The manipulation and control of vortex states in superconducting
systems are of great interest in view of possible applications,
for which mesoscopic materials are good candidates. In this work,
we studied the annihilation dynamics and the dissipative aspects of
an Abrikosov's vortex-antivortex pair in
a mesoscopic superconducting system with a concentric hole. The
generalized time-dependent Ginzburg-Landau equations were numerically solved.
The main result is the appearance of a phase slip-like line due to
the elongation of the vortex and antivortex cores.
Under specific circumstances, thermal dissipation might be associated
with a sizeable relaxation of the order parameter,
so that the energy released in the annihilation of
a vortex-antivortex pair might become detectable in
measurements of the magnetization as a function of time.

\end{abstract}


\submitto{J. Phys.: Condens. Matter}

\maketitle

\section{Introduction}\label{sec1}

The study of vortex matter is an issue of greatest interest,
since the comprehension and manipulation of the vortex
motion are very important for possible applications as
in the case of control of spins by vortices \cite{berciu,Brisbois,Lopes}.
However, unusual behaviors appear when the vortices are subjected
to an environment where the confinement effects emerge, as
is the case of mesoscopic systems. As an example,
we cite the formation of multivortex states, where one has
coexistence of single and giant vortices.
\footnote{A giant vortex is a multiquanta
vortex with a single core.} In mesoscopic systems,
this state tends to follow  the geometry of
the sample \cite{buzdin,milosevic,sardella2,
	schweigert,melnikov, geim1,palacios,misko,zhao2}.
Also, due to interactions between vortices and
the shielding currents, the formation of a giant vortex is conceivable
under certain circumstances. In this case, the cores of individual
vortices collapse into a single entity with vorticity greater
than one \cite{golubovic,baelus2,sardella3,kanda,mertelj}.

On the other hand, under specific conditions, it is possible
that a vortex-antivortex pair (V-AV) becomes stable. Such
stability has a close correspondence with the symmetry
of the system, e.g., it is possible to stabilize a state
with vorticity $3$ in a square system by $4$ vortices
located near the vertices and an antivortex in the
center of the square \cite{chibotaru}. The V-AV dynamics
was also studied in systems with holes \cite{geurts2,Sardella,zadVAV},
magnetic dots \cite{milosevic2,gomes,kramer, kapra} and
arrays of small current loops \cite{gladilin}.

The V and AV can be spontaneously formed after a quench
caused by, e.g., a hot spot \cite{shapiroPRB,shapiroEPL}.
In this scenario, as the heat is diffused and depending on
the velocity of such diffusion, the vortices are arranged
in a cluster or in a metastable ring-like configuration
\cite{shapiroPRB}. In a ring superconductor, just after
a quench, there is no interaction of the V's and the AV's
with the defects and annihilations occur. After such a
period of time, some V's and AV's leave the sample or
are trapped in the ring's hole, which generates a
magnetic flux inside the hole \cite{shapiroEPL}.

In Refs.~\cite{aranson1,aranson2}, the nucleation and
penetration of vortices were studied in very thin films
and wires under applied currents and magnetic fields.
In such cases, the vortices penetrate the samples
forming a chain in the thin films and helicoidal
lines in the wires \cite{aranson1}. It was also
shown that the normal state penetrates the superconductor
like macroscopic droplets which in the presence of defects
evolve to single vortices \cite{aranson2}.

Samples with a V-AV state were also studied by Berdiyorov
and coworkers \cite{berdiyorov}. They analyzed the
V-AV dynamics in a thin stripe with electric contacts
where a current was injected. In this system a phase
slip line is formed and the annihilation of V and AV
depends on the intensity of the applied current. This
annihilation process produces an oscillating voltage
over the contacts in a terahertz frequency \cite{berdiyorov}.
In this sense, it is interesting to mention the
work of Gulevich and Kusmartsev \cite{gulevich}
who proposed a device based on a long annular
Josephson junction where the creation, annihilation
and trapping of flux and antiflux take place. The
authors claim that their device is very sensitive and
could become a detector of microwave radiation and
magnetic fields.

In Ref.~\cite{Sardella}, Sardella and coworkers,
analyzed the annihilation of a V-AV pair in a square
 mesoscopic system with a concentric square hole.
 As a result, it was shown that, when the vortex is
 entering the system, its average velocity is of the
 order of $10^3$ m/s and, during the annihilation motion,
 due to the mutual attraction between the V and the AV,
 its average velocity reaches values of the order
 of $10^5$ m/s. Recently, Zadorosny \textit{et al.}
 \cite{zadVAV} studied similar systems and have shown that
 the V-AV pair acquires an elongated shape which creates
 a channel between the border of the system and the hole.
 In the analysis of the V-AV pair motion it was also shown
 that such specimens acquire an acceleration in the early
 and final stages of the annihilation process, with a
 nearly constant velocity motion between these stages.

In practice, mesoscopic superconducting materials have
been applied in devices like amplifiers \cite{eom},
imaging of single magnetic flux quantum (single vortex)
\cite{vasyukov}, single electron \cite{rosticher} and
single photon \cite{goltsman,kerman,salim,berdiyorov5}
detectors, and the knowledge of the V-AV dynamics in
such materials is of great importance to improve
specific characteristics to those applications.

In this work we studied the annihilation process
between a vortex and an antivortex in mesoscopic
superconducting square systems with a concentric
square hole. The study is focused on a systematic
analysis of the parameters for which the annihilation
process occurs in the superconducting region. Attention
is also given to the total energy released in such a
collision. Our results indicate that the energy
generated in such process can be associated to
frequencies in the infrared spectrum and also
that smaller systems present the higher energies.
We speculate that such system could be the heart
of a future device for detection of electromagnetic
waves in the appropriate frequency range.

This work is organized as follows. In section \ref{sec2}
we briefly delineate the theoretical formalism used
to simulate the mesoscopic systems. In section
\ref{sec3} we present the results obtained from
the simulations and, subsequently, discuss them.
In section \ref{sec4}, we present our conclusions.

\section{Theoretical Formalism}\label{sec2}

The time-dependent approach for the Ginzburg-Landau
equations, proposed by Schmid \cite{Schmid},
provides a temporal evolution of the order parameter
$\psi$ and the vector potential $\textbf{A}$ for a
superconducting material submitted to an external
applied magnetic field and/or a transport current.
Such approach is appropriate to describe most phenomena
which occur in the resistive state. For our purposes,
it will be important to use the equations for the
energy dissipated due to both the induced electrical
field and the relaxation of $\psi$ during the vortex
motion. It is worth to mention that this theoretical
framework has a satisfactory agreement with experiments
at temperatures larger than ${T} = 0.5{T}_{c}$
\cite{Schmid,Petikovic} however, qualitatively explains
the dynamics at lower temperatures. Those equations have
also been applied in studies with induced voltage
\cite{Vodolazov}, magnetoresistance \cite{Berdiyorov1,Berdiyorov2}
and the application of alternating external magnetic
fields \cite{Hernandez}. The equations proposed by Schmid were extended for
gap superconductors by Kramer and Watts-Tobin \cite{KramerTobin}.
Thus, the generalized time-dependent Ginzburg-Landau
(GTDGL) equations take the form:

\begin{eqnarray}
\frac{u}{\sqrt{1+\gamma^2|\psi|^2}}\Bigg(\frac{\partial}{\partial{t}} +
\frac{\gamma^2}{2} \frac{\partial|\psi|^2}{\partial{t}} + i\varphi\Bigg)\psi = && \nonumber \\
-(-i\mbox{\boldmath $\nabla$}-{\textbf{A}})^2\psi + \psi(1-T-|\psi|^2),&&
\label{eq:1}
\end{eqnarray}

\begin{equation}
\Bigg(\frac{\partial\textbf{A}}{{\partial{t}}} +
\mbox{\boldmath $\nabla$}\varphi\Bigg) = {\textbf{J}}_{s} -
\kappa^2\mbox{\boldmath $\nabla$}\times
\mbox{\boldmath $\nabla$}\times{\textbf{A}},
\label{eq:2}
\end{equation}
where the superconducting current density is given by:
\begin{equation}
{\textbf{J}}_{s} = {\rm Re}\left [ \bar{\psi}(-i\mbox{\boldmath $\nabla$}-{\textbf{A}})\psi\right ].
\label{eq:3}
\end{equation}
Here, the distances are in units of the coherence length
at zero temperature $\xi(0)$, the magnetic field is in units of the bulk
upper critical field $H_{c2}(0)$, the temperature is in units of
${T}_{c}$, time is in units of
${t}_{GL}(0) = \pi\hbar/8{k}_{b}{T}_{c}{u}$, the Ginzburg-Landau time, $\textbf{A}$
is in units of $H_{c2}(0) \xi(0)$, $\varphi$
is the scalar potential and is in units of $\hbar/2et_{GL}(0)$, $\kappa=\lambda(0)/\xi(0)$
is the Ginzburg-Landau parameter, where $\lambda(0)$ is the London
penetration length at zero temperature, and the order parameter is in
units of $\alpha_0T_c/\beta$, where ${\alpha}_{0}$ and ${\beta}$ are the
phenomenological Ginzburg-Landau parameters \cite{Sardella}.
The parameter ${u}$ is related to the relaxation of $\psi$
\cite{KramerBaratoff} and is very
important in studies with dissipative mechanisms; ${u}$ is extracted
from a microscopic derivation of the Ginzburg-Landau equations using
the Gor'kov approach \cite{Gorkov1968,Gorkov2}.
Frequently, ${u} = 5.79$ is adopted, as
determined by first principle in Ref.~\cite{KramerTobin}. In
such formulation the inelastic phonon-electron scattering time, $t_e$, is taken
into account and $\gamma = 2t_e\psi_0/\hbar$.
The GTDGL equations were numerically solved by using the
link-variable method \cite{Gropp,Milosevic} which ensures the gauge
invariance under the transformations
$\psi^{\prime}=\psi e^{i\chi}$,
${\bf A}^{\prime}={\bf A}+\mbox{$\boldmath \nabla$}\chi$,
$\varphi^{\prime}=\varphi-\partial\chi/\partial t$,
when they are discretized in a numerical
grid \cite{kogut}.
Therefore, for all times and positions we have chosen $\varphi^{\prime} = 0$,
since neither charges nor external currents are considered in this work.
The equation for the dissipated power energy was obtained by using the Helmholtz
free energy theorem for a superconductor in an external magnetic field
\cite{Schmid,Hernandez}. Such equation, in dimensionless form, is given by:

\begin{equation}
{W_{total}} = 2\Bigg(\frac{\partial\textbf{A}}{\partial{t}}\Bigg)^2
+ \frac{2u}{\sqrt{1+\gamma^2|\psi|^2}}
\Bigg[\Bigg(\Bigg|\frac{\partial\psi}{\partial{t}}\Bigg|\Bigg)^2+
\frac{\gamma^2}{4}\Bigg(\frac{\partial {|\psi|^2}}{\partial{t}}\Bigg)^2\Bigg].
\label{eq:4}
\end{equation}

The first term is the dissipation due to the induced electrical field,
${W}_{\textbf{A}}$, and the second one is due to the dissipation
related to the relaxation of the order parameter, ${W}_{\psi}$.
The dissipated power energy is given in units of
$H_{c2}^2(0)/[8\pi\kappa^2{t}_{GL}(0)]$.
As $W_{total}$ diffuses through the system, we couple the thermal
diffusion equation to the GTDGL ones. By using the approach
of Ref.~\cite{Vodolazov}, the dimensionless form of the thermal equation
can be written as:

\begin{eqnarray}
{C}_{eff}^\prime\frac{\partial{T}}{\partial{t}} =
	{K}_{eff}\mbox{\boldmath$\nabla$}^2{T}
	+\frac{1}{2} {W_{total}} - \eta({T} - {T}_{0}).
\label{eq:5}
\end{eqnarray}

Here, $\eta$ is the heat transfer coefficient
of the substrate, ${C}_{eff}^\prime = \pi^4/48{u}$ is
the effective heat capacity, and
${K}_{eff} = \pi^4/48{u}^2$ is the effective thermal conductivity.

In the first part of this work, as we do not take into account dissipative effects,
we set $u=1$ and $\gamma=0$ in eq.\ref{eq:1}. Such procedure is still well
accepted in the literature \cite{Berdiyorov1} due to the good qualitative
description of experimental data \cite{Baranov,Ivlev1,Ivlev2}
and facilitated computations \cite{Sardella}.
In the second part of the study, where the thermal dissipation and diffusion need
to be properly taken into account, ${u=5.79}$ and $\gamma=10$ were used.

\section{Results and Discussion}\label{sec3}

We divide the discussion into two distinct scenarios. First, we consider the temperature
constant throughout the system. And second, we take into account the heat diffusion
produced by the V-AV collision.

\begin{figure}[h!]
	\centering\vspace*{0.5cm}
	\includegraphics[scale=3.2]{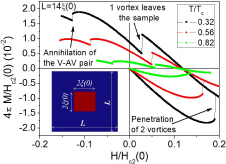}
	\caption{(Color online) Magnetization as a function of the applied magnetic
		field for the $L=14\xi(0)$ system. In such sample, two vortices are nucleated
		in the first penetration. In the decreasing field branch, one vortex is
		untrapped and leaves the sample and in the branch of negative fields the
		annihilation of the V-AV pair takes place. The inset shows an illustration
		of the simulated system.}
	\label{figone}
\end{figure}

\subsection{Constant Temperature}

The simulations of the annihilation dynamics were carried out by adopting $u=1$ and
$\kappa=5$. The value for $\kappa$ is equivalent to those of some low
critical temperature superconductors, such as the metallic alloy Pb-In \cite{poole}.
The systems were simulated with a concentric square hole of side $l=2\xi(0)$,
as shown in the inset of figure~\ref{figone}. For each system, the external
magnetic field was varied in steps of $\Delta H=10^{-3}H_{c2}(0)$ and the
temperature in steps of $\Delta T=0.2 T_c$.

\begin{figure}[h!]
  \centering\vspace*{0.5cm}
  \includegraphics[scale=3.2]{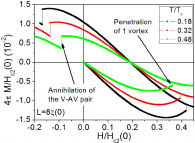}
  \caption{(Color online) Magnetization as a function of the applied
  	magnetic field for the $L=8\xi(0)$ system. In such sample, only one
  	vortex is nucleated in the first penetration and the annihilation
  	of the V-AV pair takes place in the branch of negative fields.}
  \label{figtwo}
\end{figure}

Figure~\ref{figone} shows the magnetization versus applied magnetic field,
$M(H)$ curves for the system with $L=14\xi(0)$ at different temperatures.
In this case, two vortices nucleate into the sample being trapped by the
hole. As $H$ is decreased, one vortex leaves the system and, when the
field is inverted, an AV penetrates the sample while a vortex remains
trapped in the hole. Thus, a V-AV pair is formed and each specimen moves
toward each other until their mutual annihilation. The same process
occurs in smaller systems as can be seen in figure~\ref{figtwo}
for a system with $L=8\xi(0)$. The main difference is that in the
first penetration only one vortex is nucleated and trapped by the
hole. It is interesting to note that the annihilation process was
detected even in systems with an effective superconducting region
smaller than the size of a vortex core, i.e., smaller than $2\xi(0)$.
Our simulations show that, in this case, both V and AV elongate to
accommodate themselves into the superconducting material, so that
the cores run against each other in a straight track, resembling
a phase slip line, even though the order parameter is not exactly
zero along this line. This aspect will be further discussed ahead in this paper.

\begin{figure}[h!]
  \centering\vspace*{0.5cm}
   \includegraphics[scale=3.2]{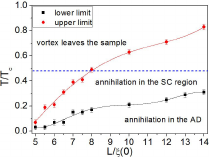}
  \caption{(Color online) The $T(L)$ diagram indicating the domains for
  	which the annihilation of a V-AV pair occurs in the superconducting
  	region. The lines which links the points are only a guide for the eyes.
  	The dashed line indicates the temperature, i.e., $T=0.48T_c$, for
  	which we analyzed the dynamics of the simulated systems.}
  \label{figthree}
\end{figure}

In order to determine the parameters, such as the range of temperatures
and the lateral sizes of the system, for which the annihilation process
takes place in the superconducting region, a $T(L)$ diagram was built
and the result is shown in figure~\ref{figthree}. As described in
this figure, below the line characterized by square
symbols, namely, the lower limit, when $H$ is inverted, the V
remains trapped in the hole. Then, an AV penetrates the system
and moves toward the center of the sample. The penetrated AV and
the trapped flux interact attractively, what causes an acceleration
of the AV, which falls in the hole and cancels the flux which was
already inside. On the other hand, above the upper limit (circles),
 when $H$ is inverted, the V is untrapped and leaves the sample before
 the nucleation of an AV. In between such lines, the annihilation of
 the V-AV pair occurs in the superconducting region.

\begin{figure}[h!]
  \centering\vspace*{0.5cm}
  \includegraphics[scale=5]{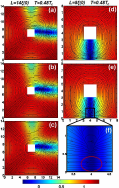}
  \caption{(Color online) Intensity of $|\psi|$ during the annihilation
  	process at $T=0.48T_c$. The black lines indicate the superconducting
  	currents flowing around the system. From panel (a) to (c), it is
  	shown the dynamics for $L=14\xi(0)$: in (a) an AV penetrates the
  	system with a V trapped in the hole; (b) the vortex leaves the
  	hole and in (c) it is shown the overlap of the currents during
  	the annihilation. From panel (d) to (f), it is shown the dynamics
  	for $L=8\xi(0)$. It is worth to note that in this case the dynamics
  	occurred near the upper threshold line shown in figure~\ref{figthree}.
  	In (d) the V moves toward the border of the system; in (e) an
  	AV starts to penetrate the system and (f) is a zoom showing
  	the distortion of the currents due to the nucleation of an AV.
  	It is also shown that the currents circumvents the vortex and/or
  	the antivortex in such a way it forms a cone-like profile.}
 \label{figfour}
\end{figure}

The horizontal line in figure~\ref{figthree} indicates the isothermal
where annihilation dynamics were analyzed. In figure~\ref{figfour} we
exhibit some images of the intensity of $|\psi|$ which summarize the
annihilation process for systems with two distinct sizes, i.e.,
$L=14\xi(0)$ $-$ figures~\ref{figfour}(a) to (c) $-$ and
$L=8\xi(0)$ $-$ (d) to (f) $-$ at $T=0.48T_c$.

\begin{figure}[h!]
  \centering\vspace*{0.5cm}
  \includegraphics[scale=0.4]{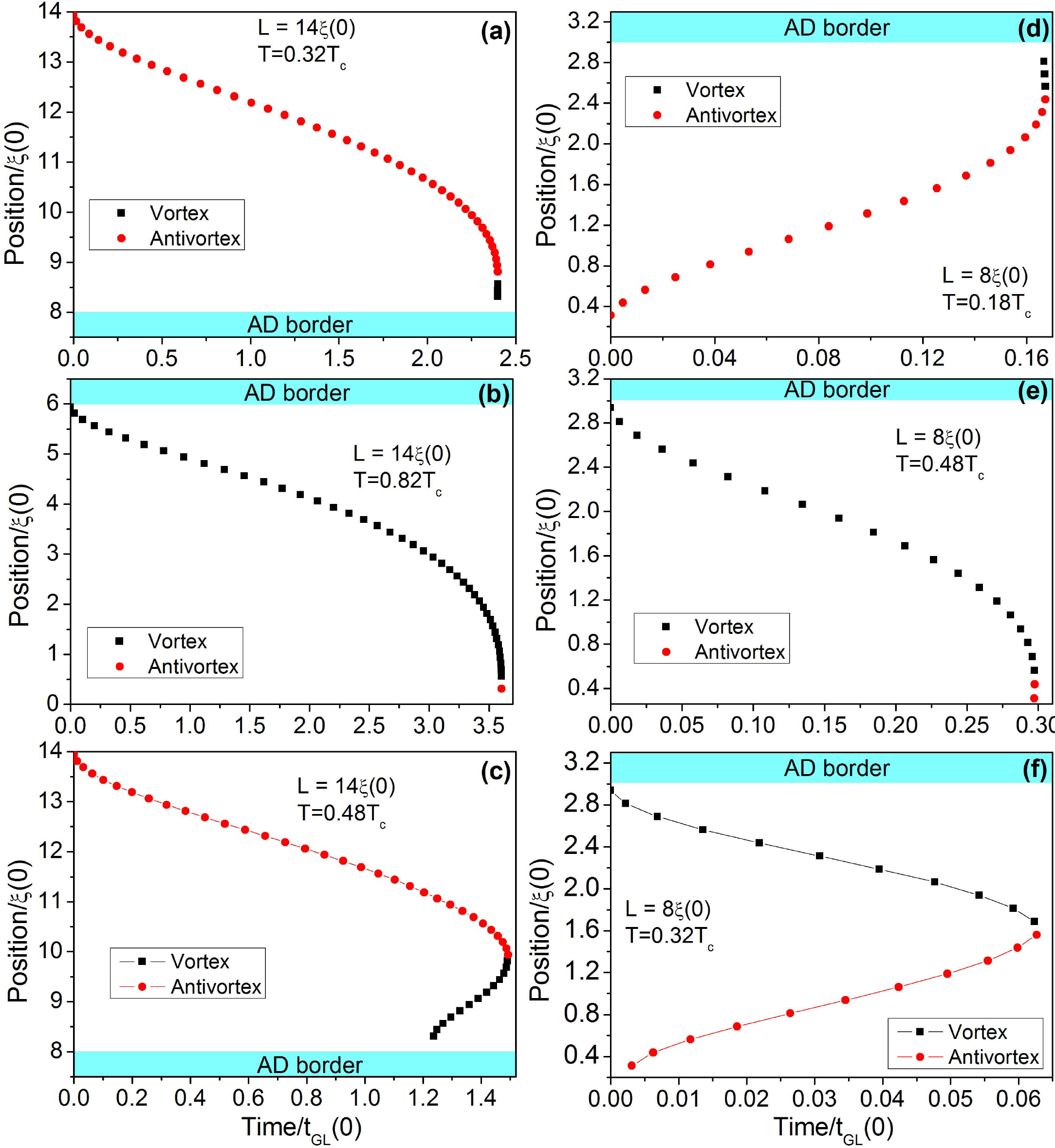}
  \caption{(Color online) Position as a function of time in
  	temperatures near the lower and upper threshold line of
  	figure~\ref{figthree}. (a) $T=0.32T_c$, (b) $T=0.82T_c$
  	for $L=14\xi(0)$ system and (d) $T=0.18T_c$, (e) $T=0.48T_c$,
  	for $L=8\xi(0)$. (c) $0.48T_c$ and (f) $0.32T_c$ show the
  	motion of the pair for an intermediate temperature for the
  	systems $L=14\xi(0)$ and $L=8\xi(0)$, respectively.}
 \label{figfive}
\end{figure}

One can notice that in both systems a quasi phase slip line is
formed \cite{footnote}. Such region appears due to the attraction
between the vortex and the antivortex which causes an elongation
of their cores \cite{zadVAV}. After the annihilation, such line
disappears. In small systems, the hole and the border of the sample
are so close that the distortion of the vortex and the antivortex,
which occurs during their encounter, is sufficient to create such
a quasi phase slip line. Figure~\ref{figfour} shows the intensity
plot of $|\psi|$ and the black lines indicate the shielding
currents. We can also notice that for both systems a visible
structure is formed by the currents, which is originated in
the drag motion of the V-AV pair. In figure~\ref{figfour}(f),
one of the borders of that system is zoomed up. We can see the
distortion of the shielding current caused by the nucleation
of an AV. To study the dynamics of the V-AV motion, in
figure~\ref{figfive} we plotted the time evolution of the position
of those specimens. The evolution of the $L=14\xi(0)$ system is
depicted in panels~\ref{figfive}(a), (b) and (c) for three
distinct temperatures: in panel (a) it is basically shown the
motion of the AV which annihilates the vortex near the hole in
the vicinity of the lower limit $T=0.32T_c$ of the phase diagram
of figure~\ref{figthree}; in panel (b) we show the vortex motion
near the upper limit line $T=0.82T_c$ where the annihilation
occurs near the border; and finally in panel (c) the motion of
an intermediate temperature is shown. The same evolution is
illustrated in panels~\ref{figfive}(d), (e) and (f) for
the $L=8\xi(0)$ system at $T=0.18T_c$, $0.48T_c$, and $0.32T_c$,
respectively.

\begin{figure}[h!]
	\centering\vspace*{0.5cm}
	\includegraphics[scale=0.5]{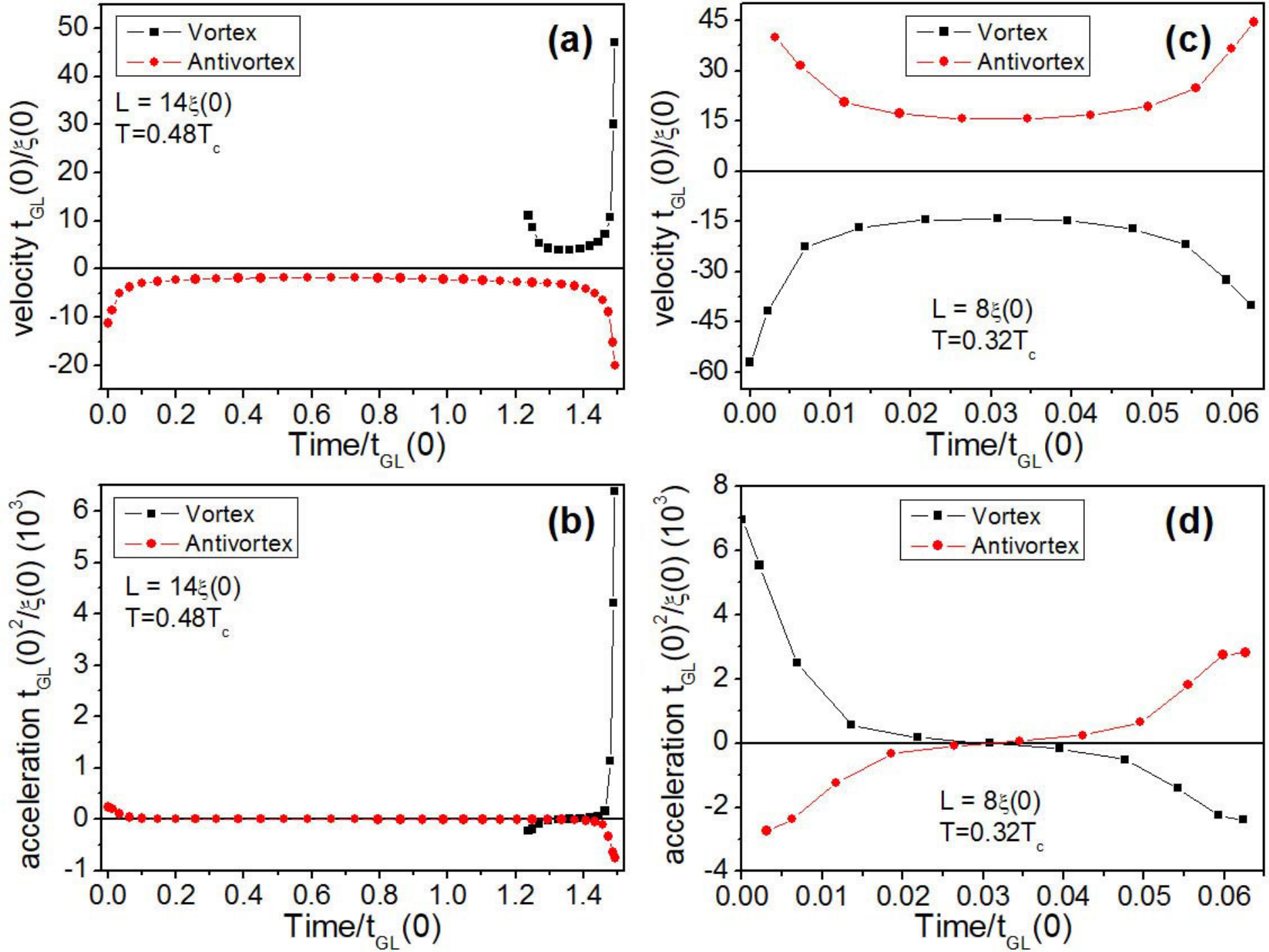}
	\caption{(Color online) Velocities and accelerations of the
		vortex and antivortex motion for the $L=14\xi(0)$ and
		$L=8\xi(0)$ systems.}
	\label{figsix}
\end{figure}

\begin{figure}[h!]
  \centering\vspace*{1cm}
  \includegraphics[scale=2]{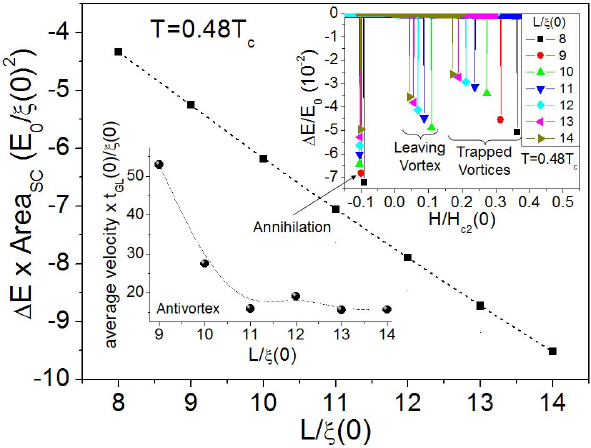}
  \caption{(Color online) The main curve shows
  		the variation of the normalized superconducting
  		energy $\times$ the normalized superconducting
  		area as a function of the lateral size of the
  		system. The $\Delta E$ was calculated as the difference
  	between the superconducting energy immediately
  	before and just after the annihilation. The energy
  	minimized in the annihilation process decreases as
  	the size of the sample is increased. The upper inset
  	shows $\Delta E$ as a function of the applied magnetic
  	field for several systems at $T=0.48T_c$. In the range
  	$0.15<H/H_{c2}<0.4$ the vortices are trapped in the
  	hole; for $0<H/H_{c2}<0.14$ one of the vortices is
  	untrapped and leaves the sample. The dips around
  	$H=-0.1H_{c2}$ occurs as the minimization of the
    energy after the annihilation. The lower
  	inset shows the average velocity as a function of the
  	lateral size of the systems. A quasi-phase slip line
  	is formed in smaller systems as a consequence of the
  	size effect. In such degraded region, the V-AV pair
  	moves faster than in larger systems where the degradation
  	does not take all the superconducting track.}
 \label{fig7b}
\end{figure}

By calculating the derivatives of the curves from
figure~\ref{figfive} (panels (c) and (f)) we obtained the
velocity and the acceleration  of the V-AV pair (see figure~\ref{figsix}).
It becomes evident that the motion of the pair is accelerated.
The amplitude of the average velocity of the AV, as an example,
decays and reach a nearly constant value as the size of the
system is increased, as shown in the lower inset of
figure~\ref{fig7b}. Further ahead this result will be discussed
in more detail.

Both the V and the AV acquire a high velocity
immediately before the annihilation, and after the collision the energy
of the system is reduced. The upper inset of figure~\ref{fig7b} shows
the variation of the superconducting energy,
$\Delta E$, as a function of $H/H_{c2}$ at $T=0.48T_c$
(dashed line of figure~\ref{figthree}). $\Delta E$ was evaluated
as the difference between the energy immediately before
and just after the annihilation. The dips in the range
\textit{(i)} $0.15<H/H_{c2}<0.4$ are due to the trapping of
a vortex (or two vortices for $L\ge10\xi(0)$) in the hole;
\textit{(ii)} for $0<H/H_{c2}<0.14$ in systems with two
penetrated vortices, one vortex leaves the sample and a
spike is detected; and \textit{(iii)} the dips around
$H=-0.1H_{c2}$ are related to the annihilation of the V-AV
pair. The main curve of figure~\ref{fig7b} shows the
values of the energy dips multiplied by the superconducting
area, $A_{SC}$, as a function of $L/\xi(0)$. Here, $A_{SC}= L^2-2^2$,
where $2^2$ is the area of the hole in
reduced units and $E_0 = \Phi_0^2/32 \pi^3\xi(0)\kappa^2$.
As we can see, the energy decreases monotonically
with $L/\xi(0)$, since the average velocity, $v_{avg}$,
for smaller systems is larger (see lower inset of the same figure).

\begin{figure}[h!]
  \centering\vspace*{1cm}
  \includegraphics[scale=2]{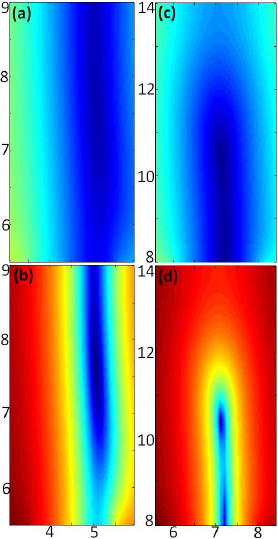}
  \caption{(Color online) Upper panels: intensity of
  	the order parameter, $|\psi|$; lower panels: for
  	a better visualizing of the degraded superconducting
  	region, we show $\log|\psi|$. In (a) and (b) we
  	show the snapshots for the  $L=9\xi(0)$ system. From
  	these panels, it can be clearly seen that no preserved
  	superconducting region remains, once the values of $v_{avg}$
    are very large. On the other hand, for the $L=14\xi(0)$
  	system is visible in both images (panels (c) and
  	(d)) that there is still a superconducting region
  	which is not entirely degraded. All snapshots were
  	taken at the instant just before the annihilation of
  	the V-AV pair.}
 \label{fig7v4}
\end{figure}

Another signature of the size effect is the decreasing
of $v_{avg}$ as $L/\xi(0)$ is increased (see the lower
inset of figure~\ref{fig7b}). In small samples, the hole
is closer to the edge. In this case, the elongation of the
vortex and the antivortex cores takes all the region
where they are moving and in such degraded region, the
pair moves faster. Panels (a) and (b) of figure~\ref{fig7v4} show the
intensity of $|\psi|$ and $\log|\psi|$ with $L=9\xi(0)$
where one can see that the superconducting state is degraded
to some degree all around the sample. On the other hand,
the degraded region for larger systems does not take all
the extension between the border and the hole. Then, the
motion of the V-AV pair is influenced by the intact
superconducting region and $v_{avg}$ decreases as
$L/\xi(0)$ increases. For $L\ge11\xi(0)$,  $v_{avg}$
becomes nearly independent of the size of the system.
In panels~\ref{fig7v4}(c) and (d), the intensity of
$|\psi|$ and $\log|\psi|$ for $L=14\xi(0)$ system present
a less degraded region which is responsible for lower
$v_{avg}$. In real units, for the Pb-In compound,
$v_{avg}$ is of the order of $10^6$m/s and such a high
velocity is due to the attraction between the V and the AV.

In the next subsection, we will analyze the kinematic
aspects of the annihilation of an Abrikosov's V-AV pair.

\subsection{Heat diffusion}\label{subsec}

\begin{figure}[h!]
  \centering\vspace*{1.5cm}
  \includegraphics[scale=3.5]{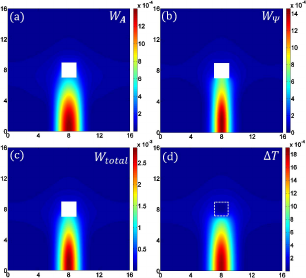}
  \caption{Intensity of the different contributions
    of dissipated power
  	energy, (a) $W_{\textbf{A}}$, (b) $W_{\psi}$ and (c) $W_{total}$,
    taken immediately before the
  	annihilation at $T=0.8 T_c$; panel (d) shows the variation of the temperature,
  	$\Delta T$, around the annihilation region.
  	The dissipation and the heat diffusion are concentrated
  	at the regions where the annihilation occurs and is
  	narrower for $W_{\psi}$ since it originates from the
    relaxation of $\psi$.}
 \label{fig8}
\end{figure}

\begin{figure}[h!]
  \centering\vspace*{0.5cm}
  \includegraphics[scale=4]{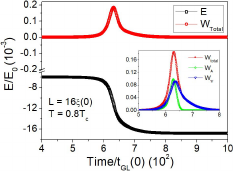}
  \caption{Superconducting ($E$) and dissipated energy
  	($W_{total}$) densities as a function of time in the
  	vicinity of the annihilation for the $L=16\xi(0)$
  	sample at $T=0.8 T_c$. The inset shows that the tail of $W_{total}$
    is related to the relaxation of $\psi$ after the annihilation,
  	as evidenced by the curves of $W_{\textbf{A}}$ and $W_{\psi}$.}
 \label{fig9}
\end{figure}

The data discussed so far were obtained without taking
into account dissipation and heat diffusion processes.
When, however, heat transfer is considered, the values of the
energy of figure~\ref{fig7b} are still valid. In
this part of the present study, we simulated similar systems as
those ones described previously. We used equations
(\ref{eq:4}) and (\ref{eq:5}) to estimate the
dissipated energy, $W_{total}$, and the variation of
the temperature, $\Delta T/T_c$, during the motion and
the annihilation of a V-AV pair. The analysis was
carried out for the $L=16\xi(0)$ sample at $T=0.8T_c$,
$\gamma=10$ and $u=5.79$.
In figure~\ref{fig8} the intensities of $W_{\textbf{A}}$, $W_{\psi}$, $W_{total}$
and $\Delta T/T_c$ are shown  in panels (a), (b), (c) and (d), respectively.
The snapshots were taken immediately
before the annihilation. As $W_{\textbf{A}}$ is related to
dissipation of normal currents, it takes a wider region in the sample
than $W_{\psi}$. On the other hand, since $W_{\psi}$
is due to the relaxation of $\psi$, a narrower region is
dominated by this dissipation mechanism. Recently, Halbertal
and coworkers have shown to be possible imaging thermal
dissipation in nanoscopic systems by using
nanoSQUIDs \cite{Zeldov}. Then, our theoretical approach
should be experimentally confirmed since $\Delta T$
is of the order of $10^{-3}T_c$ [see panel (d)], which is in the
sensitivity range of such devices.

It is worth to note in figure~\ref{fig8} that the dissipation
and the increase of the temperature are concentrated
in the annihilation's region.
Additionally, during the annihilation process, there is
no subsequent penetration of V-AV pairs.

Figure~\ref{fig9} presents $W_{total}$ and the
superconducting energy, $E$, as a function of
time during the annihilation; $W_{total}(t)$ has a tail which
is due to the different time scale of $W_{\textbf{A}}$
and $W_{\psi}$ and is associated to the relaxation of
$\psi$, as evidenced in the inset of this figure.
One can also note that both dissipation mechanisms
$W_{\textbf{A}}$ and $W_{\psi}$ have the same intensity. As
a consequence, both contributions must be taken into account for
a better description of the dissipative processes.

\begin{figure}[h!]
  \centering\vspace*{0.5cm}
  \includegraphics[scale=4.0]{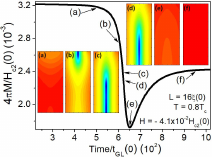}
  \caption{Magnetization as a function of time during
  	the V-AV annihilation process. The snapshots of
  	$\log|\psi|$ show the dynamics of the V-AV pair
  	motion before and after the collision.
  	(a) ($t=5t_{GL}(0)$) the vortex is trapped in
  	the hole; (b) ($t=6.06t_{GL}(0)$) the vortex leaves
  	the hole; (c) ($t=6.2t_{GL}(0)$) the antivortex
  	penetrates the sample; (d) ($t=6.23t_{GL}(0)$) the
  	V-AV collision; (e) ($t=6.51t_{GL}(0)$) degradation
  	due to the local heating and (f) ($t=9.5t_{GL}(0)$)
  	the near recovered superconducting region.}
 \label{fig10}
\end{figure}

Since the magnetization is a measurable quantity,
in figure~\ref{fig10}, the $M(t)$ curve is shown
for the $L=16\xi(0)$ system and
$T=0.8T_c$. The positive signal of $M$ is due to both
the negative applied magnetic field and the vortex
trapped in the hole. The insets show the snapshots of
$\log|\psi|$ focusing in the region where the V-AV pair
is formed. In panel (a) ($t=5t_{GL}(0)$), it is shown the state
where the V is still trapped in the hole; panel (b) ($t=6.06t_{GL}(0)$)
corresponds to the instant when the V leaves the
hole; and in panel (c) ($t=6.21t_{GL}(0)$),
	the AV penetrates the sample. At $t=6.23t_{GL}(0)$
	(panel (d)) corresponds to the instant of the V-AV collision, i.e., the
	very moment when the cores of V and AV are superimposed.
	Just after the annihilation, the temperature reaches
	its maximum value (as shown in figure~\ref{fig8}(d)).
	The local increasing of the temperature generates
	further degradation of the superconducting state of
	the surroundings (see the peak of $W_{total}$ in figure~\ref{fig9}),
	as can be seen in panel
	\ref{fig10}(e) at the instant  $t=6.51t_{GL}(0)$.
	After the annihilation, the system begins to recover the
	local superconducting state at $t=9.5t_{GL}(0)$
	(see panel (f)). Since $M$ is a response function, its
characteristic time is different from that one of the
$W_{total}$ and the inflexion of the first one
(where the annihilation occurs) does not
match with the peak of the last one. After the
annihilation, the remained response is due to the surface
of the sample. The time during which $M(t)$ changes
appreciably is $\delta t \approx 800t_{GL}(0)$, which
is of the order of nanoseconds, since $t_{GL}\approx 10^{-13}$
seconds for the Pb-In alloy \cite{poole}. Thus, to detect the
annihilation of an Abrikosov's V-AV pair using magnetometry
would require a device with resolution in the timescale of nanoseconds.

\section{Conclusions}\label{sec4}
In this work we studied the annihilation dynamics
of Abrikosov's V-AV pairs in a mesoscopic superconductor
with a central hole. In the collision process, the cores
of the V and the AV elongates and a phase slip-like
(PSlike) appears. In very small samples, $L < 10\xi(0)$,
the PSlike degrades superconductivity on the entire region
where the pair is moving increasing their average velocity.
For systems with $L > 10\xi(0)$, non degraded superconducting
regions remain in the moving area implying a lower average
velocity. The formation of a V-AV pair and its consequent
annihilation in the superconducting region depends on both
the size of the system and the temperature. Then, we
built an $L(T)$ diagram which can be used as a guide
for the predicted occurrence of the annihilation
process. The variation of the
superconducting energy just before and immediately
after the annihilation, multiplied by the
superconducting area, increases as the size of
the samples decreases, in accordance with the
average velocity of the antivortex, which is lower for larger samples.
We coupled the thermal diffusion and the dissipated
energies equations to the GTDGL ones.
Although, the dissipative
term related to the relaxation of the order parameter has
been neglected in many works, here we have evidenced that
it is a significant contribution to the dissipative processes.
We also verified that the time of the V-AV collision is
of the order of nanoseconds. Thus, conceivably,
the annihilation of Abrikosov's V-AV pairs can be
detected by measuring the magnetization response
of the sample as a function of time, what would
require fast and sensitive detection scheme.
Another interesting aspect is that the local
increase of the temperature in the annihilation
is of the order of $10^{-3}T_c$, which can be
measured by a SQUID thermal sensor as
described by Halbertal et al.\cite{Zeldov}.

\ack{We thank the Brazilian Agencies CNPq and the S\~{a}o
	Paulo Research Foundation (FAPESP), grants 2007/08072-0,
	2012/04388-0 and 2016/12390-6, for financial support.}


\section*{References}
\begin{thebibliography}{10}

\bibitem{berciu} Berciu M, Rappoport T G and Jank\'{o} B 2005 \textit{Nature} \textbf{435} 71

\bibitem{Brisbois} Brisbois J, Motta M, Avila J I, Shaw G, Devillers T, Dempsey N M, Veerapandian S K P, Colson P, Vanderheyden B,
Vanderbemden P, Ortiz W A, Nguyen N D, Kramer R B G and Silhanek R B G 2016 \textit{Sci. Rep.} \textbf{6} 27159
\bibitem{Lopes} Lopes R F, Carmo D, Colauto F, Ortiz W A, Andrade A M H, Johansen T H, Baggio-Saitovitch E and Pureur P 2017 \JAP \textbf{121} 013905
\bibitem{buzdin} Buzdin A I and Brison J P 1994 \PL \textit{A} \textbf{196} 267
\bibitem{milosevic} Milo\v{s}evi\'c M V, Kanda A, Hatsumi S, Peeters F M and Ootuka Y 2009 \PRL \textbf{103} 217003
\bibitem{sardella2} Sardella E and Brandt E H 2010 \SUST \textbf{23} 025015
\bibitem{schweigert} Schweigert V A, Peeters F M and Singha Deo P 1998 \PRL \textbf{81} 2783
\bibitem{melnikov} Melnikov A S, Nefedov I M, Ryzhov D A, Shereshevskii I A, Vinokur V M and Vysheslavtsev P P 2002 \PR \textit{B} \textbf{65} 140503(R)
\bibitem{geim1} Geim A K, Grigorieva I V, Dubonos S V, Lok J G S, Maan J C, Filippov A E and Peeters F M 1997 \textit{Nature} \textbf{390} 259
\bibitem{palacios} Palacios J J 2000 \PRL \textbf{84} 1796
\bibitem{misko} Misko V R, Xu B and Peeters F M 2007 \PRL \textbf{76} 024516
\bibitem{zhao2} Zhao H J, Misko V R, Peeters F M, Oboznov V, Dubonos S V and Grigorieva I V 2008 \PR \textit{B} \textbf{78} 104517
\bibitem{golubovic} Golubovic D S, Milo\v{s}evi\'c M V, Peeters F M and Moshchalkov V V 2005 \PR \textit{B} \textbf{71} 180502
\bibitem{baelus2} Baelus B J, Peeters F M and Schweigert V A 2001 \PR \textit{B} \textbf{63} 144517
\bibitem{sardella3} Sardella E, Lisboa-Filho P N and Malvezzi A L 2008 \PR \textit{B} \textbf{77} 104508
\bibitem{kanda} Kanda A, Baelus B J, Peeters F M, Kadowaki K and Ootuka Y 2004 \PRL \textbf{93} 257002
\bibitem{mertelj} Mertelj T and Kabanov V V 2003 \PR \textit{B} \textbf{67} 134527
\bibitem{chibotaru}	Chibotaru L F, Ceulemans A, Bruyndoncx V and Moshchalkov V V 2000 \textit{Nature} \textbf{408} 833
\bibitem{geurts2} Geurts R, Milo\v{s}evi\'c M V and Peeters F M 2009 \PR \textit{B} \textbf{79} 174508
\bibitem{Sardella} Sardella E, Lisboa-Filho P N, Silva C C S, Cabral L R E and Ortiz W A 2009 \PR \textit{B} \textbf{80} 012506
\bibitem{zadVAV} Zadorosny R, Duarte E C S, Sardella E and Ortiz W A 2014 \textit{Physica C} \textbf{503} 94
\bibitem{milosevic2} Milo\v{s}evi\'c M V, Peeters F M and Jank\'o B 2011 \SUST \textbf{24} 024001
\bibitem{gomes} Gomes A, Gonzalez E M, Gilbert D A, Milo\v{s}evi\'c M V, Kai Liu and Vicent J L 2013 \SUST \textbf{26} 085018
\bibitem{kramer} Kramer R B G, Silhanek A V, Gillijns W and Mochshalkov V V 2011 \\PR \textit{X} \textbf{1} 021004
\bibitem{kapra} Kapra A V, Misko V R, Vodolazov D Y and Peeters F M 2011 \SUST \textbf{24} 024014
\bibitem{gladilin} Gladilin V N, Tempere J, Devreese J T and Moshchalkov V V 2012 \NJP \textbf{14} 103021
\bibitem{shapiroPRB}Shapiro I, Pechenik E and Shapiro B Ya 2001 \PR \textit{B} \textbf{63} 184520
\bibitem{shapiroEPL} Ghinovker M, Shapiro B Ya and Shapiro I 2001 \textit{EPL} \textbf{53} 240
\bibitem{aranson1}Aranson I, Gitterman M and Shapiro B Ya 1995 \PR \textit{B} \textbf{51} 3092
\bibitem{aranson2}Aranson I, Shapiro B Ya and Vinokur V. 1996 \PRL \textbf{76} 142
\bibitem{berdiyorov} Berdiyorov G R, Milo\v{s}evi\'c M V and Peeters F M 2009 \PR \textit{B} \textbf{79} 184506
\bibitem{gulevich} Gulevich D R and Kusmartsev F V 2007 \NJP  \textbf{9} 59
\bibitem{eom} Eom B H, Day P K, LeDuc H G and Zmuidzinas J 2012 \textit{Nature Physics} \textbf{8} 623
\bibitem{vasyukov} Vasyukov D, Anahory Y, Embon L, Halbertal D,	Cuppens J, Neeman L, Finkler A, Segev Y, Myasoedov Y, Rappaport M L, Huber M E and Zeldov E 2013 \textit{Nature Nanotechnology} \textbf{8} 639
\bibitem{rosticher} Rosticher M, Maneval F R, Dorenbos S N, Zijlstra T, Klapwijk T M, Zwiller V, Lupascu A and Nogues G 2010 \textit{Appl.\ Phys.\ Lett.\ } \textbf{97} 183106
\bibitem{goltsman} Gol'tsman G N, Okunev O, Chulkova G, Lipatov A, Semenov A, Smirnov K, Voronov B, Dzardanov A, Williams C and Sobolewski R 2001 \textit{Appl. Phys. Lett.\ } \textbf{79} 705
\bibitem{kerman} Kerman A J , Dauler E A, Yang J K W, Rosfjord K M, Anant V, Berggren K K, Gol'tsman G N and Boronov B M 2007 \textit{Appl.\ Phys.\ Lett.\ } \textbf{90} 101110
\bibitem{salim} Salim A J, Eftekharian A and Majedi A H 2014 \JAP \textbf{115} 054514
\bibitem{berdiyorov5} Berdiyorov G R, Milo\v{s}evi\'c M V and Peeters F M 2012 \textit{Appl.\ Phys.\ Lett.\ } \textbf{100} 262603
\bibitem{Schmid} Schmid A 1966 \textit{Physik der Kondensierten Materie} \textbf{5} 302
\bibitem{Petikovic} Petkovi\'c I, Lollo A, Glazman L I and Harris J G E 2016 \textit{Nat. Commun.} \textbf{7} 13551
\bibitem{Vodolazov} Vodolazov D Y, Peeters F M, Morelle M and Moshchalkov V V 2005 \PR \textit{B} \textbf{71} 184502
\bibitem{Berdiyorov1} Berdiyorov G R, Chao X C, Peeters F M, Wang H B, Moshchalkov V V and Zhu B Y 2012 \PR \textit{B} \textbf{86} 224504
\bibitem{Berdiyorov2} Berdiyorov G R, Milo\v{s}evi\'c M V, Latimer M L, Xiao Z L, Kwok W K and Peeters F M 2012 \PRL \textbf{109} 057004
\bibitem{Hernandez} Hernandez A D and Dominguez D 2008 \PR \textit{B} \textbf{77} 224505
\bibitem{KramerTobin} Kramer L and Watts-Tobin R J 1978 \PRL \textbf{40} 1041
\bibitem{KramerBaratoff} Kramer L and Baratoff A 1977 \PRL \textbf{38} 518
\bibitem{Gorkov1968} Gor'kov L P and G M \`Eliasberg 1968 \textit{Sov. Phys. JETP} \textbf{27} 328
\bibitem{Gorkov2} Gor'kov L P and Kopnin N B 1975 \textit{Sov.\ Phys.\ Usp.\ } \textbf{18} 496
\bibitem{Gropp} Gropp W D, Kaper H G, Leaf G K, Levine D M, Palumbo M and Vinokur V M 1996 \textit{Journal of Computational Physics} \textbf{123} 254
\bibitem{Milosevic} Milo\v{s}evi\'c M V and Geurts E 2010 \textit{Physica C} \textbf{470} 791
\bibitem{kogut} Kogut J 1979 \RMP \textbf{51} 659
\bibitem{Baranov} Baranov V V, Balanov A G and Kabanov V V 2011 \PR \textit{B} \textbf{84} 094527
\bibitem{Ivlev1} Ivlev B I, Kopnin N B and Maslova L A 1980 \textit{Solid Satate Commun.} \textbf{51} 986
\bibitem{Ivlev2} Ivlev B I, Kopnin N B and Larkin L A 1985 \textit{Solid Satate Commun.} \textbf{61} 337
\bibitem{poole} Poole Jr. C P, Farach H A, Creswick R J and Prozorov R 2007 \textit{Superconductivity} 2nd ed.\ (The Netherlands: Elsevier Academic Press).
\bibitem{footnote}A phase slip line is a line along which the order parameter vanishes. This is clearly not the case in the present scenario. However, we can see that the order parameter is very small in a quite elongated region. It is in this sense that we use the denomination \textit{quasi phase slip line}
\bibitem{Zeldov} Halbertal D, Cuppens J, Shalom M Ben, Embon L, Shadmi N, Anahory Y, Naren H R, Sarkar J, Uri A, Ronen Y, Myasoedov Y, Levitov L S, Joselevich E, Geim A K and Zeldov E 2016 \textit{Nature} \textbf{539} 470
\end{thebibliography}
\end{document}